\documentclass[a4paper,11pt,onecolumn]{article}
\usepackage{latexsym}
\usepackage{graphicx}

\title{\bf Physical variability of the magnetic field of some stars }
\author{V.D. Bychkov$^{1}$, L.V. Bychkova$^{1}$, J. Madej$^{2}$  \\
$^{1}$ Special Astrophysical Observatory, Russian Academy of Sciences \\
$^{2}$ Warsaw University Observatory, Poland  }

\date{    }

\begin{document}

\begin{titlepage}
\maketitle

\begin{abstract}
Apparent variability of the longitudinal magnetic fields in
most stars is caused by rotation, which quantitavely changes projection
of the magnetic field configuration on the line of sight. This is a
purely geometrical effect and is not related to possible intrinsic
changes of the field. In some stars we observe changes of the magnetic
phase curve with time, which means that parameters of the magnetic
field change. Such changes occur in some objects in time scale of
seveal year, which is few orders of magnitude faster than predicted
by theory. Those changes imply need for improvement of the theory of
magnetic field evolution. We demonstrate changes of the rotational
phase curves in few stars.
\end{abstract}

\vspace{1.0cm}

  \begin{flushright}
    \vspace{2cm}
    {\it submitted to }  \\[2mm]
    {\rm Stars: from collapse to collapse } \\
    {\rm Special Astrophysical Observatory} \\
    {\rm 2016 October 3-7, Nizhnij Arkhyz} \\
    {\rm Karachai-Cherkessian Republic (Russia) } \\[15mm]
  \end{flushright} 
\par\vspace{3mm}

\end{titlepage}

\section{OT Ser }

This is a red dwarf star of the spectral type M1.5V and mass 
0.55 $M/M_{\odot}$ and is a well known flare star with the period
of rotation $P_{rot} = $ 3.424 days. Longitudinal magnetic field $B_e$
measurements of the star were published in the paper by Donati et al. 
(2013). Measurements of the magnetic field were carried out in two sets,
separated approximately by six months. The magnetic behavior of this
star significantly changed in this time period.

Fig. 1 shows two magnetic phase curves of OT Ser derived from $B_e$ 
measurements obtained in both sets. As can be seen from Fig. 1, magnetic
behavior of this object has significantly changed in just half a year.

It is well known that atmospheres of such stars rotate differentially 
and it causes constant generation of local magnetic fields on the surface.
Due to chaotic motions lines of force of these local fields sporadically
entwine with one another which eventually causes flares in the red dwarf
atmosphere. Flare activity in general and even annihilation of the 
strongest local magnetic fields (superflares) do not affect global
magnetic field of these stars (Bychkov et al. 2015). 

Local fields can be generated, accumulate their energy and annihilate in
presence of the global magnetic field. Probably only sometimes local magnetic
fields add and reinforce one another and finally form a global magnetic 
field of a complex structure in some approximation close to the dipole 
configuration. We believe that such a process was seen in OT Ser.
In other words, parameters of the magnetic field configuration in 
OT Ser considerably changed in time period of about six months.

\begin{figure}[ht!]
\includegraphics[width=8cm]{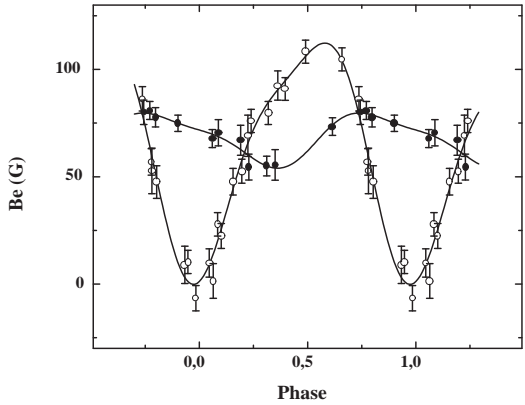}
\caption{Phase curves of OT Ser obtained from magnetic observations
    separated by half a year. }
\end{figure}

\section{ HD190073 = V1295 Aql. }
This is a well known Herbig Ae/Be star. Magnetic monitoring of this 
star was performed under the MiMeS program (Alecian et al. 2013). This
monitoring revealed that the magnetic behavior of the star significantly
changed during a very short time period (less than 4 years). It is well 
seen in Fig. 2, which was taken from Alecian et al. (2013). Global 
longitudinal magnetic field became variable with the period of rotation
$P_{rot} = 39.8 \pm 0.5 $ days. Authors of this discovery (Alecian et 
al. 2013) suggested, that the change was due to interaction between
frozen relic magnetic field and generation of new emerging field in
the convective core. Again, we observe rapid change of physical parameters
of the magnetic field configuration on the time interval from months to
years.

\begin{figure}[ht!]
{\rotatebox{90}{\includegraphics[width=6cm]{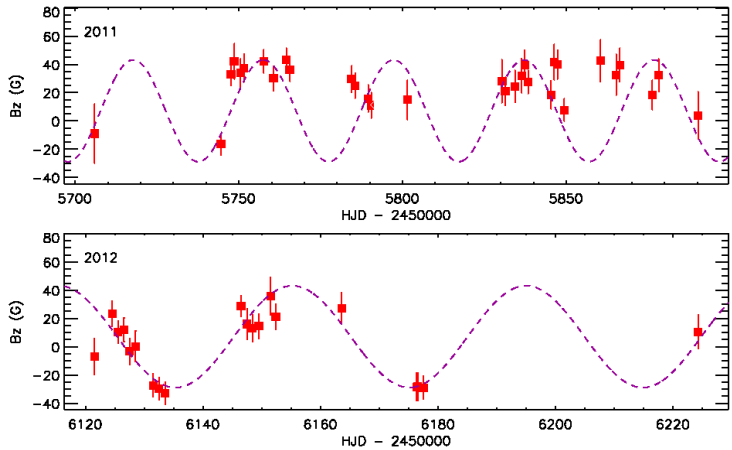}}}
\caption{Phase curve drawn with the recent magnetic measurements
obtained in two separated sets of observations. This Figure was taken
from Alecian et al. (2013). }
\label{fig:2}
\end{figure}

\section{ $\alpha^{2} CVn.$ }
This is probably the most famous magnetic chemically peculiar Ap star
showing abnormal content of Si, Cr, Eu and Hg. Magnetic field of the 
star was discovered by Babcock (1958). Since then it was actively studied
by various authors, but the most important high-precision measurements
were obtained by Wade et al. (2000) and Silvester et al. (2012). Both
research groups carried out the analysis of their observations with 
precise Least Squares Deconvolution (LSD) method defined by Donati et 
al. (1997) and Wade et al. (2000), the latter variant known as the 
WLSD method. 

The first set of high-precision magnetic field measurements by Wade 
et al. (2000) for $ \alpha^{2}$ CVn consisted of 18 data points,
obtained during 700 days (1.9 years) with the central date JD2550848.56.
Fig. 3 presents their $B_e$ points (open circles) and the average 
magnetic phase curve, the latter drawn with the average accuracy of $B_e$
points of 27.8 G (dotted line). The second set of high-precision magnetic
measurements by Silvester et al. (2012) consists of 27 measurements 
obtained during 1295 days (3.5 years) with the central date JD2554722.51.
These estimates are also displayed in Fig. 3 (filled circles) and the 
corresponding average magnetic phase curve was determined with the 
average uncertainty 9.4 G (solid line). Time interval between the 
midpoints of both sets equals 3874 days, or 10.6 years. 

Fig. 3 shows, that the average magnetic phase curve markedly changed
during this time period. Calculated differences between both phase
curves as function of the rotational phase are shown in Fig. 4, which 
also shows the total uncertainty of both phase curves, equal 29.4 G
(1$\sigma$ uncertainty). The highest differences occur at phases 
close to the peak strength of the effective magnetic field and approach 
222 G, which equals 7.5$\sigma$. At phases near the minimum $B_e$ 
differences between both phase curves amount to $4.3\sigma$. These 
are very meaningful differences. 

Observed changes of the magnetic phase curve in $ \alpha^{2} $ CVn,
if real, developed rapidly in the time period of about 10 years. This
is much less then the time scale of evolutionary changes predicted by
Krause and Raedler (1980).

Reality of the above phase curve changes in $ \alpha^{2} $ CVn is
clear when we compare with other Ap stars observed by Wade et al. 
(2000) and Silvester et al. (2012), with the same intrumentation
and LSD method used for analysing of raw polarimetric observations.
Other well known magnetic Ap stars, HD62140 and HD71866, did not show
any significant difference between magnetic phase curves which were
derived from both papers, see Bychkov et al. (2016). 

Secular changes of the rotational magnetic phase curve of $\alpha^2$
CVn still need further confirmation by new high-precision measurements
of the longitudinal field strength in this star.
Over 10 years is gone after the last set of $B_e$ points for this
star was obtained by Silvester et al. (2012). Therefore, we plan to
run a new research program on the longitudinal magnetic field observations
using the 6-meter SAO telescope after completing of new Eshel spectrometer
of high spectral resolution at Special Astrophysical Observatory
(Valyavin et al. 2014).

\begin{figure}[ht!]
\includegraphics[width=8cm]{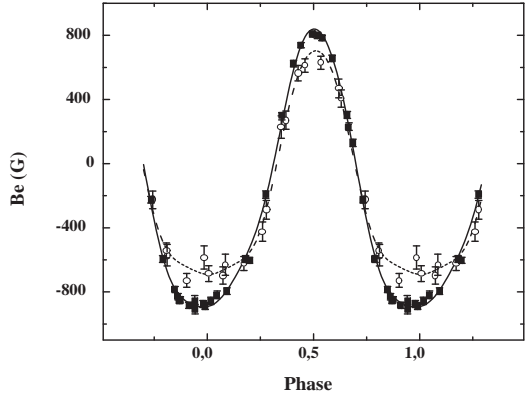}
\caption{ Empty circles denote data from Wade et al. (2000) and filled circles 
are data taken from Silvester et al. (2012). }
\label{fig:3}
\end{figure}

\begin{figure}[ht!]
\includegraphics[width=8cm]{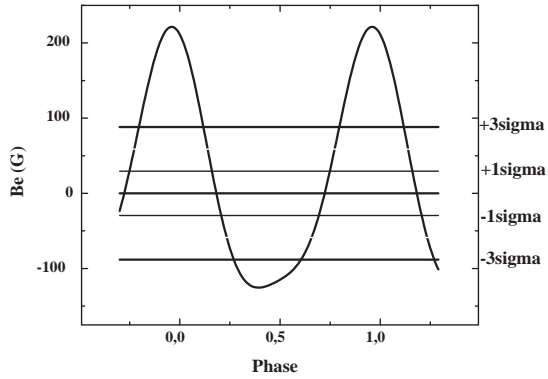}
\caption{Solid curve presents the difference between the average phase curve 
fitted to observations by Wade et al. (2000) minus the second phase
curve fitted to measurements by Silvester et al. (2012). Horizontal
lines define uncertainties of the difference curve. }
\label{fig:4}
\end{figure}

\section{ Conclusions}
Abundance spots on surfaces of Ap stars were considered as very stable
structures which could change only in very long timescales. However, as 
was shown by Kochukhov et al. (2007), chemical spots on some CP stars 
evolve in time scale of about 4 years ($\alpha$ And, HgII 3984).
Therefore, these processes are much more rapid than was previously
assumed. It would be extremely interesting to obtain distribution maps 
for various elements on the surface of $\alpha^2$ CVn in 10 - 15 years
from now and compare with the distributions presented already by Silvester
et al. (2014).

Magnetic field of a complex structure must evolve much faster than a simple
dipole (Krause \& Raedler 1980). Such complex magnetic field was observed
in $ \alpha^{2}$ CVn. Theoretical evaluations of the timescale of the
quadrupole field evolution give the scale of the order $\sim 10^4$ years
according to Krause \& Raedler (1980). 

Evolution timescale of the magnetic field was estimated as $3\times 10^7$
years from observations of stars in open clusters of different ages 
(Landstreet et al. 2008). However, this estimate was obtained for many
stars and represents only some average value. Moreover, their estimate 
in general refers to stars which have simple dipole magnetic fields. 
According to Bychkov et al. (2016), 68 \% of Ap stars have simple dipole
magnetic field and only 32 \% have more complex quadrupole fields.

We demonstrated existence of quite rapid change of the rotational phase
curve of the longitudinal magnetic field of Ap star $\alpha^2$ CVn. 
This is a signature of the intrinsic changes of the global magnetic
field parameters on the star's surface. Time scale of these variations, 
if real, is much shorter than the existing theoretical predictions
which gave estimates of much longer timescales for some average magnetic
star.

Magnetic field of this particular object needs further monitoring in
future to verify the above hypothesis on fast global field changes. 

\section{ Acknowledgements }
This work was supported by the Russian Science Foundation (project No.
14-50-00043).

\end{document}